\documentclass[12pt]{iopart}

\usepackage{iopams}
\usepackage{graphicx}           
\begin{document}

\title{Thermal Impact on Spiking Properties in Hodgkin-Huxley Neuron with Synaptic Stimulus}

\author{Shenbing Kuang}
    \address{School of Science, Wuhan University of Technology,
        Wuhan, 430070, China}
\author{Jiafu Wang}
    \address{School of Science, Wuhan University of Technology,
        Wuhan, 430070, China}
    \address{State Key Laboratory of Advanced Technology for Materials
        Synthesis and Processing, Wuhan, 430070, China}
    \ead{jasper@whut.edu.cn}
\author{Ting Zeng}
    \address{School of Science, Wuhan University of Technology,
        Wuhan, 430070, China}
\author{Aiyin Cao}
    \address{School of Science, Wuhan University of Technology,
        Wuhan, 430070, China}

\begin{abstract}
The effect of environmental temperature on neuronal spiking behaviours
is investigated by numerically simulating the temperature dependence of
spiking threshold of the Hodgkin-Huxley neuron subject to synaptic
stimulus. We find that the spiking threshold exhibits a global minimum
in a ``comfortable temperature" range where spike initiation needs
weakest synaptic strength, indicating the occurrence of optimal use of
synaptic transmission in neural system. We further explore the
biophysical origin of this phenomenon in ion channel gating kinetics
and also discuss its possible biological relevance in information
processing in neural systems.
\end{abstract}


\maketitle

Neuronal spiking is the basis of communication and coding in nervous
systems and has been studied for decades
\cite{Rabinovich2006,Feng2003,Makarov2001}. Spiking behaviours of a
neuron come from the activation and inactivation of various ion
channels, which may vary due to the influences of environmental factors
\cite{Hodgkin1952e,Schweitzer2000}. Temperature is one of the most
important regulators to neuronal activities
\cite{Vasilenko1989,Miller1994,Xu1994,Radmilovich2003}.
Electrophysiological experiments on neural spiking activities are
mostly conducted either at room temperature in vitro or at body
temperature in vivo, thus our knowledge about how a neuron behaves in a
continuously changing temperature environment is usually fragmental.
Variation in temperature leads to two main impacts in the spiking
dynamics of excitable neurons: one is on the maximum ion channel
conductances \cite{Moore1958,Cao2005} and the other is on ion channel
gating kinetics \cite{Hodgkin1952d,Bezanilla1978,Zhao2005}, hence
altering the shape and amplitude of action potentials
\cite{Hodgkin1949} and the generation
\cite{Rosenthal2002,Volgushev2000} and propagation \cite{Huxley1959} of
spikes. Neuronal spiking activities can be described by means of firing
properties, such as firing rates \cite{Rieke1997}, the precise timing
of spikes \cite{Holden2004,Chacron2004,VanRullen2005}, or the spiking
threshold behaviors \cite{Sjodin1958,Guttman1962,Yu2001,Kuang2006}. The
temperature dependence of spiking threshold of squid giant axons under
current injection has been investigated experimentally
\cite{Sjodin1958,Guttman1962,Guttman1966} and numerically
\cite{FitzHugh1962,FitzHugh1966}. More realistically, in fact, neurons
communicate with each other via synaptic connections and respond to
synaptic stimuli by firing spikes, so as to detect, transmit, and
process neural information. It is an interesting question how the
environmental temperature affects the spiking properties of a neuron
stimulated by realistic synaptic input.

This study is to approach the above topic by examining the temperature
dependence of spiking threshold of a neuron subject to synaptic
stimulus, based on widely-accepted Hodgkin-Huxley (HH) neuron model, as
an example, which was originally proposed to account for the excitable
properties of giant squid axons \cite{Hodgkin1952d} and is now regarded
to be an useful paradigm that accounts naturally for the spiking
behaviors of real neurons. Our results reveal that there is an
environmental temperature range where the spiking threshold shows a
global minimum, implying that the optimal use of synaptic transmission
occurs in neural systems. The biophysical origin of this interesting
phenomenon together with its biological relevance is also investigated
and discussed.

The dynamics of the HH model with synaptic stimulus is described by the
following coupled differential equations:
\begin{equation}
\label{Eq1}
\begin{array}{rcl}
dV/dt &=&(I_{syn}-I_{ion})/C, \\
dm/dt &=&[m_\infty (V)-m]/\tau _m(V,T), \\
dh/dt &=&[h_\infty (V)-h]/\tau _h(V,T), \\
dn/dt &=&[n_\infty (V)-n]/\tau _n(V,T),
\end{array}
\end{equation}
with $V$ being the membrane potential, $C$ the membrane capacity, $n$
the activation variable of potassium channel, and $m$ and $h$ the
activation and inactivation variables of sodium channel, respectively.
$m_\infty $, $h_\infty $, $n_\infty $ and $\tau _m$, $\tau _h$, $\tau
_n$ represent the saturated values and the time constants of the gating
variables, respectively. The $\tau$'s in Eqs.\,(\ref{Eq1}) depend on
the environmental temperature $T$, usually by being divided by a
$Q_{10}$ factor of $a$, $
Q_{10}(T,a) = a^{(T-T_0)/10}
$ with the $a$ value being usually chosen as $a = 3$ (suggested in
Ref.\,\cite{Hodgkin1952e}) and $T_0 = 6.3\,^{\circ}$C denoting the
experimental temperature for the original model construction
\cite{Hodgkin1952d} . The ionic current $I_{ion}$ includes the usual
sodium, potassium, and leak currents:
\begin{equation}
\label{Eq2}
I_{ion}=G_{Na}(T)m^3h(V-V_{Na})+G_K(T)n^4(V-V_K)+G_L(T)(V-V_L),
\end{equation}
where $V_{Na}$, $V_K$, $V_L$ are the reversal potentials for the
channel currents. The maximum channel conductances $G_{Na}$, $G_K$, and
$G_L$ are regulated by environmental temperature with a $Q_{10}$ factor
of $a = 1-1.5$ (see Ref.\,\cite{Hodgkin1952e}). The synaptic input is
modeled by $I_{syn} = g_{syn}(t)(V-V_{syn})$ with $V_{syn}$ being the
synaptic reversal potential and $g_{syn}(t)$ being the time-dependent
post-synaptic conductance, $g_{syn}(t) = G_{syn}\alpha(t-t_0)$, where
$t_0$ represents the onset time of the synapse, $G_{syn}$ determines
the peak of synaptic conductance and $\alpha(t) =
(t/\tau_{syn})\exp(1-t/\tau_{syn})$, $t>0$, with $\tau_{syn}$
determining the characteristic time of the synaptic interaction. In
this study we choose $\tau_{syn} = 2 $\,ms, and $V_{syn} = 0 $\,mV to
mimic excitatory synapse input in neural system \cite{Yu2001,Koch1999}.
The other values of parameters can be found in Ref.
\cite{Hodgkin1952d}. The spiking threshold here is characterized as the
critical value of $G_{syn}$, by which the membrane potential of the
stimulated neuron exceeds a voltage threshold $V_{th}$ (chosen as
$V_{th} = -20 $\,mV here).

\begin{figure*}
\includegraphics{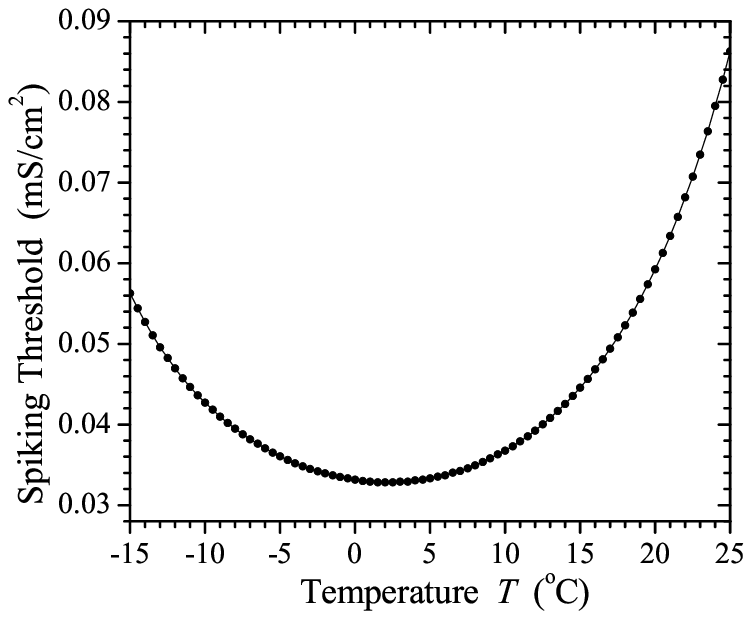}
\caption{\label{Figure1} The dependence of spiking threshold of an
HH neuron subject to synaptic stimulus on environmental temperature.
$Q_{10}$ factors of $a = 1.25$ and $a = 3$ are employed for the
maximum ion conductances (the $G$'s in Eq.\,(\ref{Eq2})) and the
gating time constants of ion channels (the $\tau$'s in
Eqs.\,(\ref{Eq1})), respectively.}
\end{figure*}

Our investigation begins with the question how environmental
temperature affects the spiking threshold of HH neuron stimulated by
synaptic input. In figure~\ref{Figure1}, one may find that the spiking
threshold undergoes a U-shaped dependence on the environmental
temperature, i.e., there is a global minimum spiking threshold in a
temperature range. We refer to this temperature range as ``comfortable
temperature'' one for the neuron. In the comfortable temperature range
the neuron needs weakest synaptic stimulus to fire spikes; from the
engineering perspective this facilitates the extraction, transmission
and processing of information in neural systems. Our result suggests
that environmental temperature can make the neurons communicate easier
with each other by maximizing the utility of synaptic transmitters. The
optimal use of synaptic stimulus in nervous system is of great
biological significance for real neurons.

\begin{figure*}
\includegraphics{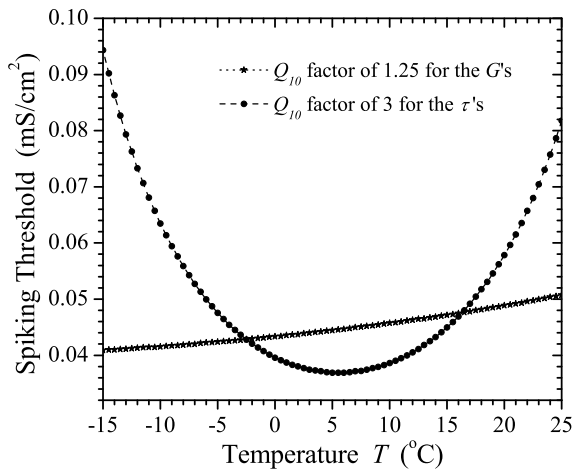}
\caption{\label{Figure2} The influence of temperature on the spiking
threshold via maximum channel conductances and via channel gating
kinetics , respectively. Dots: only the maximum ion channel
conductances (the $G$'s in Eq.\,(\ref{Eq2})) are multiplied by a
$Q_{10}$ factor of $a = 1.25$. Stars: only the time constants of
gating variables (the $\tau$'s in Eqs.\,(\ref{Eq1})) are divided by
a $Q_{10}$ factor of $a = 3$.}
\end{figure*}

Theoretically, a U-shaped dependence of a physical quantity
characteristic may result from the competition of (at least) two
contrary factors/aspects. In a neural system, environmental temperature
influences the maximum ion conductances and the channel gating rates,
as demonstrated by experimental observations \cite{Hodgkin1952e} and
explicitly included in our model simulation. To find out what factors
account for this U-shaped dependence phenomenon, we make a comparative
examination. The effects of temperature regulation on spiking threshold
through the maximum ion conductances and through the gating time
constants of channel variables (see figure~\ref{Figure2}) are
investigated respectively. If only the influence of temperature on the
maximum ion conductances is considered, the spiking threshold increases
monotonously with temperature. In direct contrast, the sole temperature
effect via gating kinetics of ion channels gives rise to a temperature
dependence of spiking threshold that almost resemble the control result
in figure~\ref{Figure1}. Thus one can reach the conclusion that it is
the dominant role of temperature on the gating kinetics of ion channels
that yields the phenomenon of optimal use of synapse transmission in
neural system, though changes in maximum channel conductances
simultaneously contribute slightly to the elevation of spiking
threshold with increasing environmental temperature.

\begin{figure*}
\includegraphics{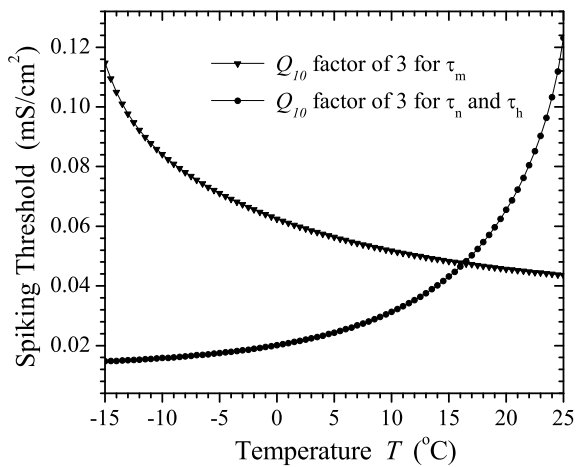}
\caption{\label{Figure3} The influence of temperature on the spiking
threshold via the thermal regulations on the activation/inactivation
variables. If only the time constant of sodium channel activation
variable $m$ is regulated with a reciprocal $Q_{10}$ factor of $a =
3$, the spiking threshold decreases with increasing temperature
(triangles); if only the time constants of sodium channel
inactivation variable $h$ and potassium channel activation variable
$n$ are divided by a $Q_{10}$ factor of $a = 3$, spiking threshold
increases with temperature (dots).}
\end{figure*}

So far we have been aware of that the occurrence of optimal use of
synaptic transmission attributes mainly to the thermal impacts on the
gating kinetics of ion channels, the direct link lacks between these
two: why does a monotonous dependence of gating time constants on
temperature result in a non-monotonous temperature dependence of
spiking threshold in neural systems? As well known, the activation of
sodium ion channel depolarizes the membrane potential and forms the
rise phase of an action potential; on the contrary, the activation of
potassium ion channel along with the inactivation of sodium ion channel
forms the decay phase of an action potential. Is it the competition of
the gating behaviours between these two types of channels that yields
the U-shaped temperature dependence of spiking threshold? The answer to
this question is unfortunately negative (simulation results are not
shown here). Thus we divide the three gating variables of ion channels
into two categories: one is the sodium ion channel activation variable
$m$, which is beneficial for a neuron to initiate spikes, and the other
includes the sodium ion channel inactivation variable $h$ and the
potassium ion channel activation variable $n$, which serve to terminate
action potentials. We speculate that it is the interaction between
these two contrary (competitive) aspects that lead to the  temperature
dependence of the neuron's spiking threshold. To confirm our
interpretation, we
 design further comparative explorations. In figure~\ref{Figure3}
one can find contrasting dependencies of spiking threshold on
environmental temperature between the thermal impact via activation
kinetic of sodium ion channel alone and that via the inactivation of
sodium ion channel together with the activation kinetic of potassium
ion channel. These results are in good agreement with our speculation.

\begin{figure}
\includegraphics{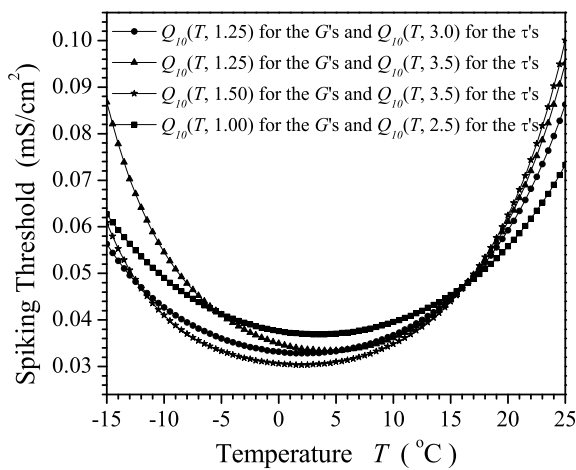}
\caption{\label{Figure4} The temperature dependencies of spiking
threshold of the HH neuron in a variety of $Q_{10}$ factors for
maximum ion conductances (the $G$'s in Eq.\,(\ref{Eq2})) and the
gating time constants of channel variables (the $\tau$'s in
Eqs.\,(\ref{Eq1})). The $Q_{10}$ factors are chosen within the
physiologically suitable range.}
\end{figure}

In the above simulation results for the temperature dependence of
spiking threshold are obtained by choosing the specific $Q_{10}$
factors of $a=3$ for time constants of channel gating variables and $a
= 1.25$ for the maximum ion conductances. Although other $Q_{10}$
factors might also be used for the neural system
\cite{Hodgkin1952e,Rothman2003}, they are not expected to yield
essential difference and the U-shaped temperature dependence still
occurs. Examples for the HH neuron are shown in figure~\ref{Figure4},
where one can see that temperature dependencies of spiking threshold
are very similar to the control case (see figure~\ref{Figure1}). The
characteristic time $\tau_{syn}$ of synaptic interaction may also vary
with environmental temperature, but again it does not lead to any
essentially different result against our conclusions. In fact, the
elegance of optimal use of synaptic transmission in a range of
environmental temperature in squid giant axons can be further
generalized to other nervous systems, such as the cochlear nucleus
neuron in auditory system \cite{Zeng}. Presumably, this subcellular
mechanism of thermal impacts on the neuronal spiking via ion channel
kinetic behaviours can serve as the base for the psychophysical
evidence that various animals live optimally in a specific range of
environmental temperature.

In summary, we have investigated the effects of environmental
temperature on the spiking behaviours of a neuron subject to synaptic
stimulus. Our results reveal that the spiking threshold shows a global
minimum in the so-called comfortable temperature range, implying the
occurrence of optimal use of synaptic transmission in neural system. We
further illustrate that the emergence of this phenomenon attributes
mainly to the combined competition of the temperature-dependent gating
kinetics of ion channel activation/inactivation variables. The optimal
use of synaptic transmission in neural system can, from an engineer's
perspective, largely facilitate the extraction, processing and
transmission of neural information in neurons.

\section*{Acknowledgements}
The authors are grateful to the financial support from the Key Project
of Chinese Ministry of Education (Grant No. 106115)

\section*{References}
\bibliographystyle{unsrt}
\bibliography{KuangArXiv}

\newpage

\end{document}